# Formal Specification Language Based IaaS Cloud Workload Regression Analysis

Sukhpal Singh
Computer Science & Engineering Department
Thapar University, Patiala, India
ssgill@thapar.edu

Inderveer Chana
Computer Science & Engineering Department
Thapar University, Patiala, India
inderveer@thapar.edu

*Abstract*— **Cloud Computing is an emerging area for accessing computing resources. In general, Cloud service providers offer services that can be clustered into three categories: SaaS, PaaS and IaaS. This paper discusses the Cloud workload analysis. The efficient Cloud workload resource mapping technique is proposed. This paper aims to provide a means of understanding and investigating IaaS Cloud workloads and the resources. In this paper, regression analysis is used to analyze the Cloud workloads and identifies the relationship between Cloud workloads and available resources. The effective organization of dynamic nature resources can be done with the help of Cloud workloads. Till Cloud workload is considered a vital talent, the Cloud resources cannot be consumed in an effective style. The proposed technique has been validated by Z Formal specification language. This approach is effective in minimizing the cost and submission burst time of Cloud workloads.**

I. INTRODUCTION

Cloud Computing is an approach of referring to the usage of different shared resources. The main services are provided by Cloud Computing are SaaS (Software as a Service), PaaS (Platform as a Service) and IaaS (Infrastructure as a Service). Cloud Computing is another way to control local server requests. It offers pay per use basis on demand policy by making cluster of Cloud resources [1] [13].

The relation between the dependent (Cloud workload) and independent (available resource) variables can be understood through regression analysis. Through this we can easily see how change in one variable affects other dependent variable. If the value of the independent variable is fixed then the average values of one or more dependent variables can be calculated [2].

The Cloud workload is an abstraction of work that instance or set of instances is going to perform. For Example: Running a web service or being a Hadoop data node is valid workloads [3].

The Z formal specification language is a core approach of first-order set theory. It supports numerous dissimilar approaches or styles of formal specification as well as those which are conventionally related with Z. The use of Z has been mainly for model oriented specification, by distinct functions on the language presented for this motive and called "schemas". Schemas are fundamentally just subsets of categorized record types [4]. The Z notation has distinct purposes, in the schema calculus and somewhere else, which provision the use of schemas as model oriented specifications of operations. The other methods of using the language which deviates from this standard and either fail to use schemas, or make supplementary inadequate use of the schema notation [5].

The main goal of our work is to establish a technique for efficient management of Cloud workloads and available resources along with workload resource matrix. The motivation behind this paper is to propose workload resource efficient mapping approach for an IaaS.

The paper is structured as follows: Related work has been presented in Section 2. In Section 3, a description of Cloud workload allocation approach has been presented. The correlation between Cloud workload and resource has been described in Section 4. The validation of proposed approach has been presented in Section 5. The conclusion and the future work have been presented in Section 6.

II. RELATED WORK

The work related to our framework has been described in this section. Qi Zhang et al. [6] apply a regression-based approximation of the CPU demand of customer transactions on available Cloud resources. They used estimation of a simple network of queues in a systematic model, every queue expressing a layer, and illustrate the approximation's effectiveness for modeling different Cloud workloads with a fluctuating transaction combination over time. They explore issues that influence the efficiency and the accuracy of the suggested performance prediction models through the use of the TPCW (Transaction Processing Performance Council) benchmark and its three different transactions combinations. The final outcomes demonstrates that regression-based approach offers an easy and controlling solution for efficient capacity scheduling and manage efficiently resources of multi-tier applications in fluctuating workload situations [6].

Mauricio Breternitz et al. [7] presented the SWAT approach (Synthetic Workload Application Toolkit) and describe the outcomes from a set of trials on specific key Cloud workloads. The design, deployment, provisioning, implementation, and data collecting of man-made compute workloads on groups of random size automates by this software platform. SWAT gathers and aggregates information from OS calls interfaces, micro architecture particular PCs (program counters) and application implementation records [8]. On the basis of collecting data, it is used to describe the special effects of network traffic, file I/O, and computation on program working. The result is examined to deliver vision into the design and deployment of Cloud workloads. Every workload is categorized according to its scalability with the amount of server nodes and Hadoop server workloads, sensitivity to network features (bandwidth, latency), and computation vs. input output amount as these values controlled via workload-specific parameters. They also explain the micro-architectural features that provide vision on the microarchitecture of processors well-matched for these Cloud workloads [8].

Soila Kavulya et al. [9] tells that a thorough study of the features of Cloud workloads executing in Map Reduce environments benefits both the Cloud providers and Cloud consumers: the Cloud provider can use this information to create improved scheduling judgments', while the Cloud consumers can study what features of their workloads influence performance. They categorize resource utilization patterns, workload patterns, and causes of failures. They used an instance-based learning method that deeds temporal locality to forecast workload end times from past figures and classify probable performance difficulties in using dataset [9].

The main role of this research is to create a relationship between Cloud workload and available resource by Cloud workload allocation approach and then verifying their consistency using Z specification language.

### III. CLOUD WORKLOAD ALLOCATION APPROACH

#### A. Resource – Cloud Workload Mapping Procedure

Describe the procedure of mapping of Cloud workloads with available resources according to mapping rules. The efficient mapping is created by workload analyzer according to Cloud customer and provider constraints. The Figure 1 describes the Cloud Workload – Resource Mapping Procedure.

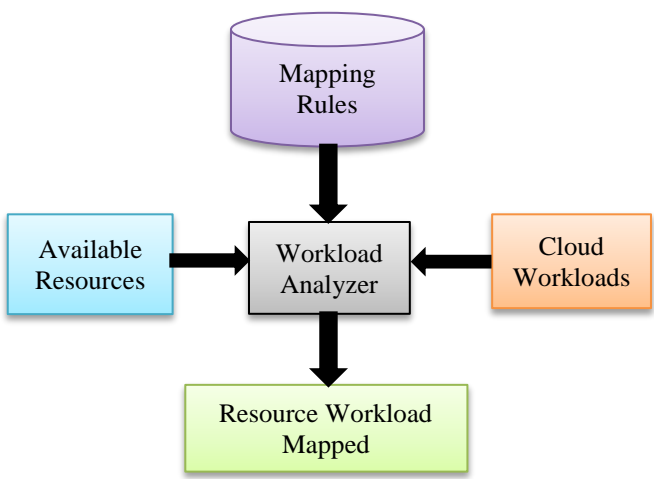

Figure 1. Resource – Cloud Workload Mapping Procedure

#### B. Cloud workload resource matrix

Rows and columns are used to describe the efficient mapping of Cloud workload and available resource. The Table I shows the Cloud workload resource matrix.

TABLE I. CLOUD WORKLOAD RESOURCE MATRIX

|    | W1 | W2 | W3 | W4 | W5 | W6 | W7 |
|----|----|----|----|----|----|----|----|
| R1 |    | √  |    |    |    |    |    |
| R2 |    |    | √  |    |    |    |    |
| R3 |    |    |    |    | √  |    |    |
| R4 | √  |    |    |    |    |    |    |
| R5 |    |    |    | √  |    |    |    |
| R6 |    |    |    |    |    | √  |    |
| R7 |    |    |    |    |    |    | √  |

### IV. CORRELATION BETWEEN CLOUD WORKLOAD AND RESOURCE

The simple linear model of regression analysis can help you to determine a correlation between Cloud workloads and resources [10]. Suppose we reckon that some variable of interest, r, is 'driven by' some other variable w. In this w is Cloud workload and r is an available resource. We then call r the dependent variable and w the independent variable. Further, assume that the association between r and w is mainly linear, but is inaccurate: moreover its determination by w, r has a random component, c, which we call the 'disturbance' or 'error'. Let $a$ index the observations on the data pairs (w, r). The simple linear model formalizes the ideas just stated:

$$r_a = \mu_0 + \mu_1 w_a + c_a$$

The $\mu_0$ and $\mu_1$ are mapping parameters. $\mu_0$ and $\mu_1$ represent the r-intercept and the slope of the relationship, respectively. In order to work with this model we need to make some assumptions about the behaviour of the error term. The basis of assumptions made regarding the error behaviour make possible to: 1) Develop measures of reliability for regression coefficients and 2) Test hypothesis about the association between w and r: draw inferences [4]. For now we'll assume three things:

$F(c_a) = 0$         c has a mean of zero for all a

$F(c_a^2) = \sigma_c^2$        it has the similar change for all a

$F(c_a, c_b) = 0$  a≠b        no correlation across observations

With the help of the values of $\mu_0$ and $\mu_1$ the workloads and resources are mapped efficiently. We have just made a bunch of assumptions about what is 'really going on' between r and w, but we would like to put numbers on the parameters $\mu_0$ and $\mu_1$. Well, suppose we are able to gather a sample of data on w and r. The task of estimation is then to come up with coefficients—numbers that we can calculate from the data, call them $\mu'_0$ and $\mu'_1$ —which serve as estimates of the unknown mapping parameters. If we can do this somehow, the estimated equation will have the form:

$$r'_a = \mu'_0 + \mu'_1 w$$

We express the expected fault or residual related with every pair of data values as the actual $r_a$ value minus the prediction based on $w_a$ along with the estimated coefficients:

$$c'_a = r_a - r'_a = r_a - (\mu'_0 + \mu'_1 w_a)$$

In a scatter diagram of r against w, this is the vertical distance between observed $r_a$ and the 'fitted value', $r'_a$, as shown in Figure 2.

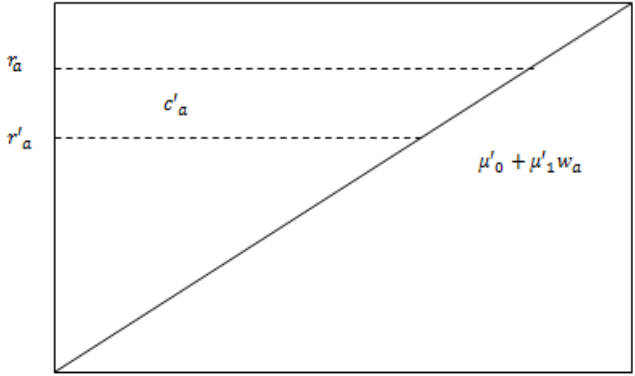

Figure 2. Cloud Workload Regression Residual

Note that we are using a different symbol for this estimated error ($c'_a$) as opposed to the 'true' disturbance or error term defined above ($c_a$). These two will coincide only if $\mu'_0$ and $\mu'_1$ happen to be exact estimates of the regression parameters $\mu_0$ and $\mu_1$. The most common technique for determining the coefficients $\mu'_0$ and $\mu'_1$ is Ordinary Least Squares (OLS) [10]: values for $\mu'_0$ and $\mu'_1$ are chosen so as to minimize the Sum of the Squared Residuals (SSR) [11]. The SSR may be written as

$$SSR = \sum c'^2_a = \sum (r_a - r'_a)^2 = \sum (r_a - \mu'_0 - \mu'_1 w_a)^2$$

It should be understood throughout that $\sum$ denotes the summation $\sum_{a=1}^{n}$, where n is the number of interpretations in the trial. The reduction of SSR is a calculus exercise: we need to find the partial derivatives of SSR with respect to both $\mu'_0$ and $\mu'_1$ and set them equal to zero. This generates two equations (known as the 'normal equations' of least squares) in the two unknowns, $\mu'_0$ and $\mu'_1$. These equations are then solved jointly to yield the estimated coefficients. We start out from:

$$\delta SSR / \delta \mu'_0 = -2 \sum (r_a - \mu'_0 - \mu'_1 w_a) = 0 \quad (1)$$

$$\delta SSR / \delta \mu'_1 = -2 \sum w_a (r_a - \mu'_0 - \mu'_1 w_a) = 0 \quad (2)$$

Equation (1) implies that

$$\sum r_a - n\mu'_0 - \mu'_1 \sum w_a = 0$$

$$\Rightarrow \mu'_0 = r' - \mu'_1 w' \quad (3)$$

while equation (2) implies that

$$\sum w_a r_a - \mu'_0 \sum w_a - \mu'_1 \sum w_a^2 = 0 \quad (4)$$

We can now substitute for $\mu'_0$ in equation (4), using (3). This yields

$$\sum w_a r_a - (r' - \mu'_1 w') \sum w_a - \mu'_1 \sum w_a^2 = 0$$

$$\sum w_a r_a - r' \sum w_a - \mu'_1 (\sum w_a^2 - w' \sum w_a) = 0$$

$$\mu'_1 = \frac{\sum w_a r_a - r' \sum w_a}{\sum w_a^2 - w' \sum w_a} \quad (5)$$

Equations (3) and (4) can now be used to generate the regression coefficients. First use (5) to find $\mu'_1$, then use (3) to find $\mu'_0$. Goodness of fit: The OLS technique ensures that we find the values of $\mu'_0$ and $\mu'_1$, which 'fit the sample data best', in the specific sense of minimizing the sum of squared residuals [11].

## V. VALIDATION OF CLOUD WORKLOAD ALLOCATION APPROACH

The confidence of correctness can be increased by augmenting the development process with formal verification, i.e., regression verification [10]. Regression verification applies formal verification techniques to continuously check development revisions in order to identify regressions early [11]. Regression verification outputs intermediate results (Correlation between Cloud Workload and Resource) in order to enable a more efficient re-verification of a revised Cloud Workload Allocation Approach relying on the very same verification process [12]. Formal specification can serve as a single, reliable reference point for who investigate Cloud workloads; map the available resources to Cloud workloads and those who verify the results [4]. In Z specification [4, 5], schemas are used to describe both the static and dynamic aspects of a system. Z decomposes specifications into manageably sized module's called schemas: Schemas are divided into three parts: 1. A state, 2. A collection of state variables and their values and 3. Operations that can change its state [5]. This section explains how the framework deals with the resources and Cloud workloads. The set of all resource names and Cloud workloads are the basic types of the specifications [5].

[RESOURCENAME, CLOUDWORKLOAD]

The first aspect of the workload analyzer is its state space.

```
┌─ WorkloadAllocationApproach ─────────
│ availableresource : ℙ RESOURCENAME
│ allocation : RESOURCENAME ↦ CLOUDWORKLOAD
├──────────────────────────────────────
│ availableresource = dom allocation
└──────────────────────────────────────
```

In our work, the space of workload analysis has been described and the two variables represent important observations which can make of the state [5].

- *availableresource* is the set of available resources.
- *allocation* is a function that when applied to certain resources (Res), create a mapping of workloads with resources associated with them.
- set *item* is the same as the domain of the function allocate the resources to which it can be validly applied.

availableresource = {Res1, Res2, Res3}

allocation = { Res1 → Cloudworkload3

Res2 → Cloudworkload2

Res3 → Cloudworkload1 }

The invariant is satisfied because allocation details a CLOUDWORKLOD for tree RESOURCENAME in allocation. There are some operations that can apply on the workload analyzer:

The first of all there is to add a new resource, and we describe it with schema:

```
┌─ AddResourceForAllocation ──────────
│ Δ WorkloadAllocationApproach
│ workload? : CLOUDWORKLOAD
│ resource? : RESOURCENAME
├──────────────────────────────────────
│ resource? ∉ availableresource
│ allocation' = allocation ∪ {resource? ↦ workload?}
└──────────────────────────────────────
```

The Δ *WorkloadAllocationApproach* alerts us to the fact that the schema is describing a state change: it introduces four variable availableresource, allocation, availableresource' and allocation'. The first two are observations of the state earlier the modification, and the most recent two are interpretations of the state after the change [5]. We expect that the set of resource known to *WorkloadAllocationApproach* will augmented with new resource.

allocation' = allocation ∪ {resource ?}

We can prove this from the specification of *AddResourceForAllocation* using the invariants on the state before and after.

availableresource ' = dom allocation '

= dom (allocation ∪ {resource? → workload?})

= dom allocation ∪ dom {resource? → workload?}

= dom allocation ∪ {resource?}

= allocation ∪ {resource?}

In *FindResourceForAllocation*, find the resources to map the Cloud workloads based on user requirements.

```
┌─ FindResourceForAllocation ─────────
│ Ξ WorkloadAllocatioApproach
│ workload! : CLOUDWORKLOAD
│ resource? : RESOURCENAME
├──────────────────────────────────────
│ resource? ∈ availableresource
│ workload! = allocation (resource?)
└──────────────────────────────────────
```

The declaration Ξ*WorkloadAllocationApproach* indicates that this is an operation in which the state does not change, the value of avilableresource' and allocation' of the observations after the operation are equal to these values availableresource and allocation. Including Ξ*WorkloadAllocationApproach* above the line has the same effect as including Δ*WorkloadAllocationApproach* above the line and two equations below it.

availableresource' = availableresource
allocation' = allocation

The other notation (!) for an output the *FindResourceForAllocation* operations take an avilableresource as input and yield corresponding mapping as output.

The most useful operation on workload analyzer is one to find which workload map with available resource. The operation has an input rank? And one output, item! Which is set of resources for allocation? There may be zero, one or more workloads map with particular resource, to whom resource item should be sent.

```
┌─ Map ───────────────────────────────
│ Ξ WorkloadAllocationApproach
│ rank? : CLOUDWORKLOAD
│ item! : ℙ RESOURCENAME
├──────────────────────────────────────
│ item! = { n : availableresource | allocation (n) = rank? }
└──────────────────────────────────────
```

This time there is no pre-condition. The item! is specified to be equal to the set of all values n drawn from the set item such that the value of the allocation function at n is rank? [5]. In general, q is a member of the set {p: L |……..p………}
exactly if q is a member of L and the condition ……q……,
obtained by replacing p with q, is satisfied:

$$q \in \{p: L | \ldots p \ldots\} \leftrightarrow q \in L \land (\ldots q \ldots)$$

$$s \in \{n: item \land allocation(s) = rank?\}$$

$$\leftrightarrow s \in item \land allocation(s) = rank?$$

A name s is in output set item! exactly if it is known to the workload analyzer and the allocation recorded for it is rank? [5].

The given below schema identify the initial state of the workload analyzer:

```
─── InitWorkloadAllocationApproach ───
WorkloadAllocationApproach
─────────────────────────────
availableresource = ∅
```

This schema describes a *WorkloadAllocationApproach* in which the set known is empty: in consequence, the function allocation empty too.

We shall add an extra output! to each action in workload analyzer. After successful execution of given process the outcome will be OK [5], but it may take the other value AlreadyMapped and NotMapped when error is detected. REPORT defines the set contains three values.

REPORT::=OK/ AlreadyMapped /NotMapped

The result should be OK after proper execution of success schema without saying how the state changes.

```
─── Success ───
output! : REPORT
─────────────
output! = OK
```

WorkloadAllocationApproach ∧ Success

The conjunction operator ∧ of the schema calculates allows us to combine this description with our previous description of *WorkloadAllocationApproach*:

The process for accurate input has defines both acts as described by *WorkloadAllocationApproach* and produces the result OK.

Schema specified that the report *AlreadyMapped* should be produced when input resource? Is already a member of mapped.

```
─── AlreadyMapped ───
Ξ WorkloadAllocationApproach
resource? : RESOURCENAME
output! : REPORT
──────────────────────
resource? ∈ Mapped
output! : already_Mapped
```

ΞWorkloadAllocationApproach specifies that if the error occurs, the state of the workload analyzer should not change.

$$RWorkloadAllocationApproach \stackrel{\wedge}{=} (WorkloadAllocationApproach) \lor AlreadyMapped$$

```
─── RWorkloadAllocationApproach ───
Δ WorkloadAllocationApproach
workload? : CLOUDWORKLOAD
resource? : RESOURCENAME
output! : REPORT
──────────────────────────
(resource? ∈ Mapped ∧
allocation' = allocation ∪ {resource? ↦ workload?} ∧
output! = OK) ∨
resource? ∈ Mapped ∧ allocation' = allocation ∧
output! : already_Mapped
```

```
─── NotMapped ───
Ξ WorkloadAllocationApproach
resource? : RESOURCENAME
output! : REPORT
──────────────────────
resource? ∉ Mapped
output! : not_Mapped
```

$$RFindResourceForAllocation \stackrel{\wedge}{=} (FindResourceForAllocation \land Success) \lor NotMapped$$

Mapped operation can be called at any time, it never results an error robust version need only add reporting of success.

$$RMapped \stackrel{\wedge}{=} Mapped \land Success$$

## VI. CONCLUSION AND FUTURE SCOPE

Using the proposed technique, IaaS resources and workloads can be efficiently managed. This paper discussed the Cloud workload regression analysis. This paper provides an approach, through which the IaaS Cloud workloads and the resources can be understood easily. Regression analysis is used to analyze the Cloud workloads and identifies the relationship between Cloud workloads and available resources. The proposed technique has been validated by Z Formal specification language through different schemas. The mapping of Cloud workloads and resources can be done in an efficient manner without wastage of time and cost, the relations between Cloud workloads and resources will be easily verified through z specification language. In the Future, through this approach the Cloud workloads will be identified and categorized properly based on some QoS requirements of each and every workload, further, the characteristics and constraints for each Cloud workload will be identified. The metrics based on key Quality of Service (QoS) requirements will be identified for each Cloud workload. Based on this, the efficient scheduling of resources can be achieved through the clustering of workloads according to the key QoS requirements that will be effective in minimizing the cost and submission burst time of Cloud workloads.